\documentclass[onecolumn,amsmath,nofootinbib]{revtex4} % For astro-ph
\usepackage{graphicx}
\usepackage{url}
\begin{document}
\input{epsf.sty}

\newcommand\AJ[3]{~Astron. J.{\bf ~#1}, #2~(#3)}
\newcommand\APJ[3]{~Astrophys. J.{\bf ~#1}, #2~ (#3)}
\newcommand\APJL[3]{~Astrophys. J. Lett. {\bf ~#1}, L#2~(#3)}
\newcommand\APP[3]{~Astropart. Phys. {\bf ~#1}, #2~(#3)}
\newcommand\CQG[3]{~Class. Quant. Grav.{\bf ~#1}, #2~(#3)}
\newcommand\JETPL[3]{~JETP. Lett.{\bf ~#1}, #2~(#3)}
\newcommand\MNRAS[3]{~Mon. Not. R. Astron. Soc.{\bf ~#1}, #2~(#3)}
\newcommand\MPLA[3]{~Mod. Phys. Lett. A{\bf ~#1}, #2~(#3)}
\newcommand\NAT[3]{~Nature{\bf ~#1}, #2~(#3)}
\newcommand\NPB[3]{~Nucl. Phys. B{\bf ~#1}, #2~(#3)}
\newcommand\PLB[3]{~Phys. Lett. B{\bf ~#1}, #2~(#3)}
\newcommand\PR[3]{~Phys. Rev.{\bf ~#1}, #2~(#3)}
\newcommand\PRL[3]{~Phys. Rev. Lett.{\bf ~#1}, #2~(#3)}
\newcommand\PRD[3]{~Phys. Rev. D{\bf ~#1}, #2~(#3)}
\newcommand\PROG[3]{~Prog. Theor. Phys.{\bf ~#1}, #2~(#3)}
\newcommand\PRPT[3]{~Phys.Rept.{\bf ~#1}, #2~(#3)}
\newcommand\RMP[3]{~Rev. Mod. Phys.{\bf ~#1}, #2~(#3)}
\newcommand\SCI[3]{~Science{\bf ~#1}, #2~(#3)}
\newcommand\SAL[3]{~Sov. Astron. Lett{\bf ~#1}, #2~(#3)}

\def\affilmrk#1{$^{#1}$}
\def\affilmk#1#2{$^{#1}$#2;}

\def\affilmrk#1{$^{#1}$}
\def\affilmk#1#2{$^{#1}$#2;}
\def\be{\begin{equation}}
\def\ee{\end{equation}}
\def\bea{\begin{eqnarray}}
\def\eea{\end{eqnarray}}

\title{Cosmological neutrino mass limit and the dynamics of dark energy
 }
\author{ Jun-Qing Xia, Gong-Bo Zhao and Xinmin Zhang }

\affiliation{Institute of High Energy Physics, Chinese Academy of
Science, P.O. Box 918-4, Beijing 100049, P. R. China}

\date{\today}

\begin{abstract}

We investigate the correlation between the neutrino mass limit and
dark energy with time evolving equation of state. Parameterizing
dark energy as $w=w_0+w_1*z/(1+z)$, we make a global fit using
Markov Chain Monte Carlo technique to determine $w_0, w_1$, neutrino
mass as well as other cosmological parameters simultaneously. We pay
particular attention to the correlation between neutrino mass
$\Sigma m_{\nu}$ and $w_1$ using current cosmological observations
as well as the future simulated datasets such as PLANCK, SNAP and
LAMOST.

\end{abstract}


\maketitle

%%%%%%%%%%%%%%%%%%%%%%%%%%%%%%%%%%%%%%%%%%%%%%
%%%%%%%%%%%%%%%%%%%%%%%%%%%%%%%%%%%%%%%%%%%%%%

\setcounter{footnote}{0}

\section{Introduction}
With the accumulation of high quality observational data of cosmic
microwave background (CMB) \cite{Spergel:2006hy,wmap3:2006}, large
scale structure (LSS) of galaxies
\cite{Cole:2005sx,Tegmark:2003uf,Tegmark:2003ud} and Supernovae Type
Ia (SN Ia) \cite{Tonry03,Riess04,Riess05}, we have been pacing into
a new era of precision cosmology. In 1998, the analysis of the
luminosity-redshift relation of SN Ia revealed the current
acceleration of our universe \cite{Riess98,Perl99}. This
acceleration attributes to a new form of energy, dubbed dark energy
(DE), with negative pressure and almost not clustering whose nature
remains unveiled. The simplest candidate of dark energy is the
cosmological constant (CC) however it suffers from the fine tunning
and coincidence problems \cite{SW89,ZWS99}. To ameliorate these
dilemmas some dynamical dark energy models such as quintessence
\cite{quint}, phantom \cite{phantom}, k-essence \cite{kessence} and
quintom whose equation of state (EOS) can cross -1 during evolution
\cite{quintom} have been proposed and studied thoroughly both
theoretically and phenomenologically in the literature
\cite{quintom,Feng:2004ff,xia,Zhao:2005vj,Xia:2005ge,
Xia:2006cr,Zhao:2006bt,Xia:2006rr,Li:2005fm,Zhang:2005eg,Zhang:2006ck,
Guo:2004fq,Copeland:2006wr,relevnt}.

Neutrino physics, especially the mass limit, is another challenge
of modern science. The combined analysis of all currently
available data of neutrino oscillation experiments implies the
crucial mass differences in the neutrino mass hierarchy, say,
$\Delta m_{12}^2 \simeq 7 \times 10^{-5}$eV$^2$, $\Delta m_{23}^2
\simeq 2.6 \times 10^{-3}$eV$^2$ for solar neutrino  and  the
atmospheric neutrino mass difference respectively
\cite{Aliani:2003ns}. In the scenario of hierarchical neutrino
masses these results suggest  $m_1 \sim 0$, $m_2 \sim \Delta
m_{\rm solar}$, and $m_3 \sim \Delta m_{\rm atmospheric}$. For the
inverted hierarchy  $m_3 \sim 0$, $m_2 \sim \Delta m_{\rm
atmospheric}$, and $m_1 \sim \Delta m_{\rm atmospheric}$. However,
if the neutrino masses are degenerate, one finds $m_1 \sim m_2
\sim m_3 \gg \Delta m_{\rm atmospheric}$. These oscillation
experiments can merely constrain the squared mass differences,
$\Delta m^2$, they cannot detect the absolute value of neutrino
masses which is of great significance.

However, cosmological observations can put upper limits of the
absolute neutrino mass. For background evolution, neutrino masses
contribute to the cosmic energy budget so that modify the epoch of
matter-radiation equality, angular diameter distance to the last
scattering surface (LSS) and other related physical quantities.
For the perturbation, neutrinos become non-relativistic at late
time thus they damp the perturbation within their free streaming
scale. This suppressed the matter power spectrum by roughly
$\Delta P/P \sim - 8 \Omega_\nu/\Omega_m$ \cite{Hu:1997mj}. Dark
energy can also affect the evolution of background and
perturbation, which can mimic the behavior of neutrino to some
extent, so that there exists an obvious correlation between
equation of state of dark energy and neutrino mass.

For simplicity, Ref.\cite{Hannestad:2005gj,Goobar:2006xz}
parameterizes dark energy by constant equation of state and finds
it strongly correlates with neutrino mass. In this paper, we study
this correlation with dynamical dark energy models which are more
generic with current cosmological observations as well as with
future simulated data. Further we have determined the parameters
of dynamical dark energy and neutrino mass limit simultaneously.

The rest part of this paper is structured as follows: in the next
section we present our fitting method and data we use. In section
III we give our results. We make conclusion and discussion in the
last section.

\section{Method and data}

We choose the commonly used parametrization of the EOS of dark
energy as \cite{Linder:2002et}:
\begin{equation}\label{linder}
    w(z)=w_0+w_1\frac{z}{1+z}~,
\end{equation}
where $z$ is the redshift. Clearly the cosmological constant
corresponds to $w_0=-1,~w_1=0$ thus $\Lambda$CDM model is
incorporated in this parametrization. For neutrino mass, we use
the parameter $f_\nu$ which is defined as dark matter neutrino
fraction:
\begin{equation}\label{fnu}
    f_\nu =\rho_{\nu}/\rho_{DM}=\frac{\Sigma
    m_{\nu}}{93.105~eV~\Omega_{c}h^{2}}~,
\end{equation}
where  $\rho_{\nu}$ and $\rho_{DM}$ denote the energy density of
neutrino and dark matter respectively, $\Sigma m_{\nu}$ is the sum
of neutrino mass and $\Omega_{c}h^{2}$ is the physical cold dark
matter densities relative to critical density.

The fitting and statistics strategy we adopt is based on the
Markov Chain Monte Carlo package \texttt{CosmoMC}
\cite{Lewis:2002ah}\footnote{Available at
http://cosmologist.info/cosmomc/}, which has been modified to
implement dark energy perturbations with EOS getting across $-1$
\cite{Zhao:2005vj}. Our most general parameter space is
\be\label{para}
    \textbf{p}\equiv(\omega_{b}, \omega_{c}, \Theta_S, \tau, w_0, w_1,  f_\nu, n_{s}, \log[10^{10}A_{s}])~,
\ee where $\omega_{b}=\Omega_{b}h^{2}$ and
$\omega_{c}=\Omega_{c}h^{2}$ are the physical baryon and cold dark
matter densities relative to critical density, $\Theta_S$ is the
ratio (multiplied by 100) of the sound horizon to the angular
diameter distance at decoupling, $\tau$ is the optical depth,
$A_{s}$ is defined as the amplitude of initial power spectrum and
$n_{s}$ measures the spectral index. Assuming a \textbf{flat}
Universe motivated by inflation and basing on the Bayesian
analysis, we vary the 9 parameters above and fit to the
observational data with the MCMC method. We take the weak priors
as: $\tau<0.8, 0.5<n_{s}<1.5, -3<w_{0}<3, -5<w_{1}<5~\footnote{We
set the prior of $w_0$ and $w_1$ broad enough to ensure the EoS
can evolve in the whole parameter space.}, 0<f_{\nu}<0.5$ and a
cosmic age tophat prior as 10 Gyr$<t_{0}<$20 Gyr. Furthermore, we
make use of the HST measurement of the Hubble parameter $H_0 =
100h ~\text{km s}^{-1} \text{Mpc}^{-1}$ \cite{freedman} by
multiplying the likelihood by a Gaussian likelihood function
centered around $h=0.72$ and with a standard deviation $\sigma =
0.08$. We impose a weak Gaussian prior on the baryon density
$\Omega_b h^2 = 0.022 \pm 0.002$ (1$\sigma$) from Big Bang
nucleosynthesis \cite{bbn}.

In our global fitting we compute the total likelihood to be the
products of separate likelihoods of CMB, SNIa, LSS and
Heidelberg-Moscow experiment (HM) \cite{Kl04,Kl06,fogli2}.
Alternatively defining $\chi^2 = -2 \log {\bf \cal{L}}$, so \be
\chi^2_{total} =
\chi^2_{CMB}+\chi^2_{SNIa}+\chi^2_{LSS}+\chi^2_{HM}~. \ee For CMB
data we use the three-year WMAP (WMAP3) Temperature-Temperature (TT)
and Temperature-Polarization (TE) power spectrum with the routine
for computing the likelihood supplied by the WMAP team
\cite{WMAP3IE}. The supernova data we use are the ``gold" set of 157
SNIa published by Riess $et$ $al$ in \cite{Riess04}. We have
marginalized over the nuisance parameter \cite{DiPietro:2002cz} in
the calculation of SN Ia likelihood. For LSS information, we have
used the 3-D matter power spectrum of SDSS \cite{Tegmark:2003uf} and
2dFGRS \cite{Cole:2005sx}, Lyman-$\alpha$ forest data (Ly$\alpha$)
from SDSS \cite{lya} and recent measurement of the baryon acoustic
oscillation feature in the 2-point correlation function of SDSS
\cite{Eisenstein2005}. To be conservative but more robust, we only
use the first 14 bins of the SDSS 3-D matter power spectrum, which
are well within the linear regime \cite{sdssfit}. For Ly$\alpha$
likelihood, we modify the interpolating code\footnote{Available at
http://www.cita.utoronto.ca/~pmcdonal/LyaF/public.lyafchisq.tar.gz}
to incorporate dynamical dark energy models. For BAO likelihood, we
use the constraint\footnote{In this work we mainly focus on the
correlation between dark energy parameters and the neutrino mass
rather than the absolute value of the neutrino mass. What we are
interested in is the possible effect of BAO measurement on this
correlation. Since the BAO measurement seems not very consistent
with the other observations, such as CMB and SN Ia, on the
constraints of dark energy parameters, we don't use the full power
spectrum analysis of BAO in our calculations for simplicity.
Similarly the author of Ref.\cite{Cirelli:2006kt} also use the
parameter A to study the neutrino mass limit.}
\cite{Eisenstein2005}: \be
 A \equiv D_V(0.35)
{\sqrt{\Omega_m H_0^2}\over 0.35 c} = 0.469 \pm 0.017~,\ee \be
D_V(z) = \left[ D_M(z)^2 {cz\over H(z)}\right]^{1/3}~, \ee where
$H(z)$ is the Hubble parameter, $c$ is the speed of light and
$D_M(z)$ is the comoving angular diameter distance at a specific
redshift $z$. Thus
 \be\label{BAO}
    \chi^{2}_{BAO}=\frac{(A-0.469)^{2}}{0.017^{2}}~.
\ee Moreover, the Heidelberg-Moscow experiment uses the half time of
$0\nu2\beta$ decay to constrain the effective Majorana mass and this
translates to the constraints of sum of neutrino mass under some
assumptions \cite{DeLaMacorra:2006tu}: \be\label{HMP} \Sigma
m_{\nu}\sim 1.8\pm0.6eV~(2\sigma)~.\ee Given the Heidelberg-Moscow
experiment is controversial for the time being and seems not
consistent with other observational data, we just make a tentative
fit choosing the HM prior to illustrate the effect on the
correlation between the dark energy parameters and the neutrino mass
when the total neutrino mass is very large.

To get robust conclusion, we have also used the simulated data from
future observations of PLANCK, LAMOST and SNAP. The fiducial model
we choose is the best-fit from the WMAP3 dataset
\cite{Spergel:2006hy}: $\Omega_{b}=0.04$, $\Omega_{c}=0.20$,
$h_{0}=0.73$, $\tau=0.088$, $n_s=0.95$ and $A_s=0.68$. And we assume
the sum of neutrino mass $\sum m_{\nu}=0$ and the equation of state
of dark energy $w_0=-1$, $w_1=0$.

For the CMB simulation we consider a simple full-sky ($f_{sky}=1$)
simulation at Planck-like sensitivity and ignore the lensing
effect and the tensor information. We neglect foregrounds and
assume the isotropic noise with variance
$N_{l}^{TT}=N_{l}^{EE}/2=3\times10^{-4}\mu K^2$ (Pessimistic
Planck-like sensitivity) and a symmetric Gaussian beam of 7
arcminutes full-width half-maximum (FWHM) \cite{Lewis:2005tp}. We
use the simulated $\tilde C_{l}$ up to $l=2500$ for temperature
and $l=1500$ for polarization. The effective $\chi^2$ is: \be
\chi_{eff}^2 \equiv -2 \log \mathcal{L}=\sum_l (2l+1) \left \{
\log \left ( \frac{C_{l}^{TT}C_{l}^{EE}-(C_{l}^{TE})^2}{\tilde
C_{l}^{TT}\tilde C_{l}^{EE}-(\tilde C_{l}^{TE})^2} \right ) +
\frac{\tilde C_{l}^{TT}C_{l}^{EE}+C_{l}^{TT}\tilde
C_{l}^{EE}-2\tilde
C_{l}^{TE}C_{l}^{TE}}{C_{l}^{TT}C_{l}^{EE}-(C_{l}^{TE})^2} -2
\right \}~, \ee where $C_{l}^{XY}$ denote theoretical power
spectra and $\tilde C_{l}^{XY}$ denote the power spectra from the
simulated data. The likelihood has been normalized with respect to
the maximum likelihood, where $C_{l}^{XY}=\tilde C_{l}^{XY}$
\cite{Easther:2004vq,Perotto:2006rj}.

For the future LSS survey we consider LAMOST project. The Large
Sky Area Multi-Object Fiber Spectroscopic Telescope (LAMOST)
project as one of the National Major Scientific Projects
undertaken by the Chinese Academy of Science, aims to measure
$\sim 10^7$ galaxies with mean redshift $z \sim 0.2$
\cite{lamost}. In the measurements of large scale matter power
spectrum of galaxies there are generally two statistical errors:
sample variance and shot noise. The uncertainty due to statistical
effects, averaged over a radial bin $\Delta k$ in Fourier space,
is \cite{9304022} \be \label{eqn:dPK} (\frac{\sigma_P}{P})^2  =
2\times \frac{(2 \pi)^3}{V}\times \frac{1}{4 \pi k^2 \Delta
k}\times (1+ \frac{1}{\bar{n}P})^2~. \ee The initial factor of 2
is due to the real property of the density field, $V$ is the
survey volume and $\bar{n}$ is the mean galaxy density. In our
simulations for simplicity and to be conservative, we use only the
linear matter power spectrum up to $k \sim 0.1$ $h$ Mpc$^{-1}$.

The projected satellite SNAP\footnote{SNAP is one of the several
candidates emission concepts for the Joint Dark Energy
Mission(JDEM).} (Supernova / Acceleration Probe)would be a space
based telescope with a one square degree field of view with 1
billion pixels. It aims to increase the discovery rate for SNIa to
about 2,000 per year \cite{snap}. The simulated SN Ia data
distribution is taken from Refs.\cite{kim,Li:2005zd}. As for the
error, we follow Ref.\cite{kim} which takes the magnitude dispersion
$0.15$ and the systematic error $\sigma_{sys}(z)=0.02\times z/1.7$,
and the whole error for each data is \be
\sigma_{mag}(z_i)=\sqrt{\sigma_{sys}^2(z_i)+\frac{0.15^2}{n_i}}~,
\ee where $n_i$ is the number of supernova in the $i$'th redshift
bin.

For each regular calculation, we run 6 independent chains
comprising of 150,000-300,000 chain elements and spend thousands
of CPU hours to calculate on a cluster. The average acceptance
rate is about 40\%. And for the convergence test typically we get
the chains satisfy the Gelman and Rubin \cite{GR92} criteria where
R-1$<$0.1.

Despite our ignorance of the nature of dark energy, it is more
natural to consider the DE fluctuation whether DE is regarded as
scalar field or fluid rather than simply switching it off. The
conservation law of energy reads: \be\label{law}
    T^{\mu\nu}{}_{;\mu}=0~,
\ee where $T^{\mu\nu}$ is the energy-momentum tensor of dark
energy and ``;" denotes the covariant differentiation. Working in
the conformal Newtonian gauge, Equation(\ref{law}) leads to the
perturbation equations of dark energy as follows \cite{ma}: \bea
    \dot\delta&=&-(1+w)(\theta-3\dot{\Phi})
    -3\mathcal{H}(\delta p/\delta\rho-w)\delta~, \label{dotdelta}\\
\dot\theta&=&-\mathcal{H}(1-3w)\theta-\frac{\dot{w}}{1+w}\theta
    +k^{2}(\frac{\delta p/\delta\rho}{{1+w}}\delta+ \Psi)~ .\label{dottheta}
\eea  For the models where the EOS doesn't cross -1,  the above
equation(\ref{dotdelta}), (\ref{dottheta}) is well defined. For
the crossing models, the perturbation equation (\ref{dotdelta}),
(\ref{dottheta}) is seemingly divergent. However basing on the
realistic two-field-quintom model as well as the single field case
with a high derivative term \cite{Zhao:2005vj}, the perturbation
of DE is shown to be continuous when the EOS gets across -1, thus
we introduce a small positive parameter $\epsilon$ to divide the
full range of the allowed value of the EOS $w$ into three parts:
1) $ w> -1 + \epsilon$; 2) $-1 + \epsilon \geq w \geq-1 -
\epsilon$; and 3) $w < -1 -\epsilon $.

For the regions 1) and 3) the perturbation is well defined by
solving Eqs.(\ref{dotdelta}), (\ref{dottheta}) as shown above. For
the case 2), the perturbation of energy density $\delta$ and
divergence of velocity, $\theta$, and the time derivatives of
$\delta$ and $\theta$ are finite and continuous for the realistic
quintom dark energy models. However for the perturbations with the
above parameterizations clearly there exists some divergence. To
eliminate the divergence typically one needs to base on the
multi-component DE models which result in the non-practical
parameter-doubling. A simple way out is to match the perturbation
in region 2) to the regions 1) and 3) at the boundary and set
\cite{Zhao:2005vj,Xia:2005ge,Xia:2006cr}
\begin{equation}\label{dotx}
  \dot{\delta}=0 ~~,~~\dot{\theta}=0~.
\end{equation}
We have numerically checked the error in the range  $|\Delta w =
\epsilon |<10^{-5}$ and found it less than 0.001$\%$ to the exact
multi-field quintom model. Therefore our matching strategy is a
perfect approximation to calculate the perturbation consistently
for crossing models. For more details of this method we refer the
readers to our previous companion papers
\cite{Zhao:2005vj,Xia:2005ge,Xia:2006cr}.

\section{Results}
\begin{table*}
TABLE 1. Median values and 1$\sigma$ constrains on dark energy
parameters and sum of neutrino mass for models discussed in the
text. From left to right, we use WMAP three year data (WMAP3),
WMAP3+Riess 157 ``Gold" sample+2dF+SDSS(14bands)(LSS),
WMAP3+LSS+Lyman-$\alpha$ forest(Ly$\alpha$),
WMAP3+LSS+Lyman-$\alpha$+Baryon Acoustic Oscillation (BAO)
information, WMAP3+LSS+Ly$\alpha$+BAO+Heidelberg-Moscow experiment
data respectively. The right most column we use the simulated data
for future SNAP+PLANCK+LAMOST. For one-tailed distributed
parameter such as neutrino mass, we quote the 95$\%$ C.L. limit.

\begin{center}
\begin{tabular}{|c|c|c|c|c|c|c|}
  \hline
  \hline
&\multicolumn{6}{c|}{ $m_\nu$+Dynamical Dark Energy with Dark Energy Perturbation} \\
\hline &WMAP3 &+SN+LSS
&+Ly$\alpha$&+BAO&+Heidelberg-Moscow&Planck+SNAP+LAMOST
\\
\hline
$w_0$  &$-1.004^{+0.662}_{-0.717}$&$-1.212^{+0.228}_{-0.209}$ &$-1.138^{+0.196}_{-0.204}$&$-0.820^{+0.219}_{-0.214}$&$-0.725^{+0.277}_{-0.252}$&$-0.984^{+0.070}_{-0.069}$ \\
$w_1$  &$-1.939^{+2.459}_{-2.098}$&$0.418^{+0.823}_{-0.848}$ &$0.125^{+0.859}_{-0.842}$&$-0.924^{+1.077}_{-1.095}$&$-1.894^{+1.472}_{-1.651}$&$-0.173^{+0.266}_{-0.278}$ \\
$\Sigma m_{\nu} (eV)$  &$<4.532  $&$<2.896  $ &$<1.149  $&$<0.490  $&$0.971^{+0.225}_{-0.239}$&$<0.323  $ \\
$cov(w_0,m_{\nu})$  &$-0.067$&$-0.382 $&$-0.221$&0.110&0.086&0.0005 \\
$cov(w_1,m_{\nu})$    &$-0.077$&$-0.109$ &$-0.117$&$-0.171$&$-0.170$&$-0.295$  \\
\hline \hline
&\multicolumn{6}{c|}{$m_\nu$+Dynamical Dark Energy without Dark Energy Perturbation} \\
\hline &WMAP3 &+SN+LSS
&+Ly$\alpha$&+BAO&+Heidelberg-Moscow&Planck+SNAP+LAMOST
\\
\hline
$w_0$  &$-0.859^{+0.564}_{-0.558}$&$-1.174^{+0.181}_{-0.180}$ &$-1.170^{+0.173}_{-0.170}$&$-0.901^{+0.147}_{-0.155}$&$-0.846^{+0.180}_{-0.179}$&$-$ \\
$w_1$  &$-0.571^{+1.342}_{-1.365}$&$0.393^{+0.555}_{-0.553}$ &$0.440^{+0.567}_{-0.560}$&$-0.342^{+0.595}_{-0.585}$&$-0.960^{+0.868}_{-0.915}$&$-$ \\
$\Sigma m_{\nu} (eV)$  &$<4.684  $&$<2.786  $ &$<0.991  $&$<0.433  $&$0.899^{+0.247}_{-0.244}$&$-  $ \\
$cov(w_0,m_{\nu})$  &0.041& $-0.447$ &$-0.295$ &0.013&0.089&$-$ \\
$cov(w_1,m_{\nu})$    &$-0.143$& $ -0.017$&$0.023$ &$-0.131$&$-0.236$&$-$  \\
\hline \hline
&\multicolumn{6}{c|}{$m_\nu$+WCDM with Dark Energy Perturbation} \\
\hline &WMAP3 &+SN+LSS
&+Ly$\alpha$&+BAO&+Heidelberg-Moscow&Planck+SNAP+LAMOST
\\
\hline
$w$  &$-$&$-$ &$-$&$-0.978^{+0.071}_{-0.066}$&$-$&$-$ \\
$\Sigma m_{\nu} (eV)$  &$-  $&$-  $ &$-  $&$<0.398  $&$-$&$-  $ \\
$cov(w,m_{\nu})$  &$-$&$ -$ &$-$ &$-0.137$&$-$&$-$ \\
\hline \hline
&\multicolumn{6}{c|}{$m_\nu$+$\Lambda$CDM } \\
\hline &WMAP3 &+SN+LSS
&+Ly$\alpha$&+BAO&+Heidelberg-Moscow&Planck+SNAP+LAMOST
\\
\hline
$\Sigma m_{\nu} (eV)$  &$-  $&$- $ &$<0.299 $&$- $&$-$&$-  $ \\
\hline \hline
\end{tabular}
\end{center}
\end{table*}

In this section we present our global fitting results of dark
energy parameters $w_0,~w_1$ and neutrino mass limit. We pay
particular attention to the correlation between $w_1$, which
delineates the time evolving of dark energy, and neutrino mass.
Our main result was summarized in Table I.  We choose the general
model of 9 free parameters as illustrated in Eq.(\ref{para}) to
study the correlation between dynamical dark energy and neutrino
mass and discuss the model with DE with constant equation of state
(WCDM) and $\Lambda$CDM model for comparison. Here we only list
the parameters of dark energy and neutrino mass. Since dark energy
perturbation is crucial in determination of cosmological
parameters especially for dark energy parameters
\cite{Spergel:2006hy,Zhao:2005vj,Xia:2005ge,Zhao:2006bt}, we
include the full dark energy perturbation in our global fitting as
well as showing the result by incorrectly switching off dark
energy perturbations to reinforce this key point. We use different
data combination discussed in previous section. The cov(X,Y) is
the correlation coefficient of samples which quantifies the
correlation between the two parameters X and Y defined
as\cite{corr}: \be
cov(X,Y)=\frac{\sum^{n}_{i=1}(X_{i}-\bar{X_{i}})(Y_{i}-\bar{Y_{i}})}
{\sqrt{\sum^{n}_{i=1}(X_{i}-\bar{X_{i}})^{2}}\sqrt{\sum^{n}_{i=1}(Y_{i}-\bar{Y_{i}})^{2}}}~,
\ee where the bar denotes the mean value of parameters. If
cov(X,Y)$>$0 means X,Y is positively correlated and \emph{vice
versa}.

\begin{figure}[htbp]
\begin{center}
\includegraphics[scale=0.6]{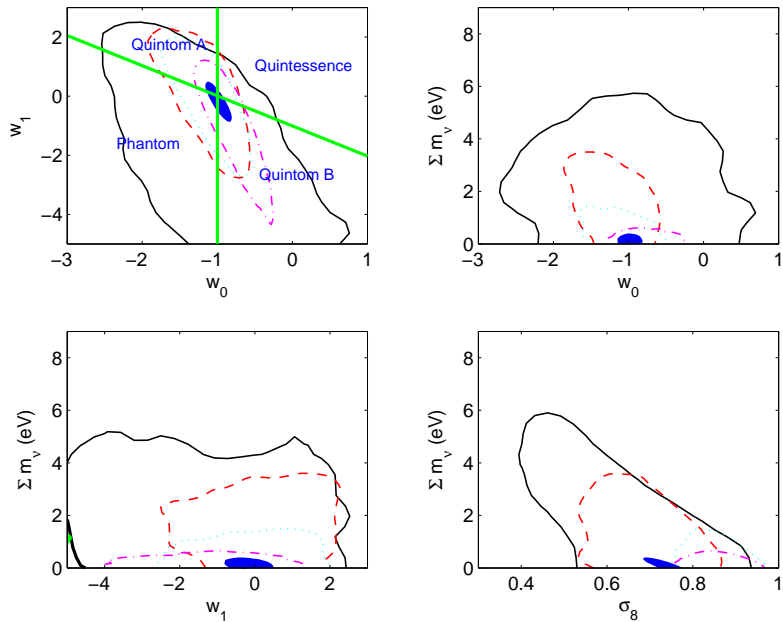}
\caption{ $95\%$ C.L.contour plots of the sum of neutrino mass
$\Sigma m_{\nu}$ with dark energy parameters $w_0,~w_1$ as well as
$\sigma_8$ with different combination of datasets. From outside in
, we use WMAP3 only (black), WMAP3+RIESS+SDSS+2dF (red),
WMAP3+RIESS+SDSS+2dF+Ly$\alpha$ (cyan),
WMAP3+RIESS+SDSS+2dF+Ly$\alpha$+BAO (purple) and simulated data of
SNAP+PLANCK+LAMOST (innermost blue shaded). The two lines in the
$w_0,~w_1$ plane distinguish the dark energy models (see text) and
their intersecting point denotes the $\Lambda$CDM model. In the
fitting we include the full dark energy perturbations of dynamical
dark energy models. \label{2d}}
\end{center}
\end{figure}

From the comparison of results with/without dark energy
perturbation, we find dark energy perturbation modifies the mean
value of all the parameters as well as enlarging the corresponding
errors which is consistence with previous analysis in the
literature
\cite{Spergel:2006hy,Zhao:2005vj,Xia:2005ge,Zhao:2006bt}. We will
merely discuss the results with dark energy perturbation
hereafter. We have seen that WMAP3 only put weak constraints on
$w_0,~w_1$ and neutrino mass since CMB data is sensitive to the
effective equation of state of dark energy defined as
\cite{Wang:1999fa}: \be\label{weff}
    w_{eff}\equiv\frac{\int da \Omega(a) w(a)}{\int da \Omega(a)}~,
\ee thus it's hard to constrain $w_0,~w_1$ and neutrino mass by
CMB data alone. Adding SN Ia and LSS data we find the mean value
of each parameter gets modified and the error bars shrink a lot.
However it is noteworthy that adding BAO data enlarges the error
of $w_0,~w_1$ while tightens the neutrino mass limit to a great
extent. If we take the HM prior in Eq.(\ref{HMP}), we find the
error of $w_0,~w_1$ become greater while the 1-D posterior
distribution of neutrino mass turns to be two-tailed. We have also
done the global fitting using the simulated data of future
observation and found the errors shrink greatly.

%\begin{figure}[htbp]
%\begin{center}
%\includegraphics[scale=0.6]{1D.eps}
%\caption{1-D posterior distribution of dark energy parameters,
%neutrino mass and $\sigma_8$. Black solid line:
%WMAP3+RIESS+SDSS+2dF+Lyman-$\alpha$+BAO +Heidelberg-Moscow (All+HM);
%Red dashed: WMAP3+RIESS+SDSS+2dF+Lyman-$\alpha$+BAO (ALL without
%HM)}
%\end{center}
%\end{figure}

\begin{figure}[htbp]
\begin{center}
\includegraphics[scale=0.6]{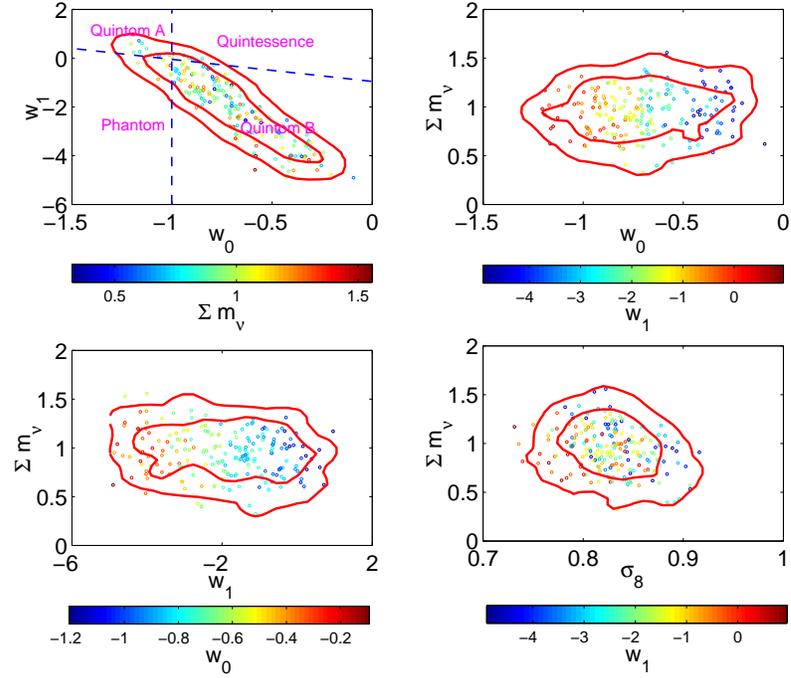}
\caption{ 3-D scatter and $68\%$ and $95\%$ C.L. contour plots of
the sum of neutrino mass $\Sigma m_{\nu}$ with dark energy
parameters $w_0,~w_1$ as well as $\sigma_8$ using
WMAP3+RIESS+SDSS+2dF+Lyman-$\alpha$+BAO +Heidelberg-Moscow. Dark
energy perturbation is also included. \label{3d}}
\end{center}
\end{figure}

\begin{figure}[htbp]
\begin{center}
\includegraphics[scale=1.2]{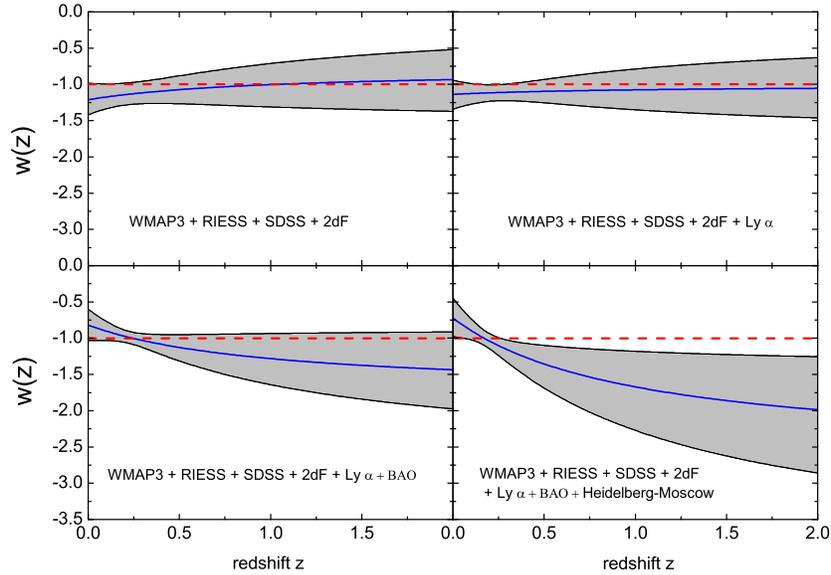}
\caption{Evolution of the equation of state of dark energy
constrained by different combination of datasets as illustrated in
the plots. The blue solid curve and the gray shaded bands show the
mean value and 1-$\sigma$ error respectively. Dark energy
perturbation is fully included.
 \label{wz}}
\end{center}
\end{figure}

In Fig.(\ref{2d}) we show the $95\%$ C.L. 2-D contours of
$w_0,~w_1$, neutrino mass and $\sigma_{8}$. From outside in, we
add data as illustrated in the figure caption. Using simulated
data we obtain the inner most shaded regions. The contours shrink
significantly while interesting correlations remain. The results
including HM prior are illustrated in Fig.(\ref{3d}). We also cut
the $w_0-w_1$ plane into four parts for different dark energy
models as we did in \cite{Xia:2005ge,Xia:2006cr,Zhao:2006bt}. The
EOS of Quintom A crossing -1 from upside down, \emph{i.e.} the EOS
is greater than -1 in the early epoch yet negative than -1
currently while quintom B crosses -1 from the opposite direction.
We find that the $\Lambda$CDM model cannot be ruled out by all the
data combination at $2\sigma$ level however the quintom-like
dynamical dark energy model is mildly favored. The one dimensional
constraints of EOS by different combination of datasets are
illustrated in Fig.(\ref{wz}). We find that the best fit models
are always allowed crossing the boundary $w=-1$ in these four
combinations. When adding the dataset of BAO measurement or HM
measurement the direction of EOS crossing the boundary will be
changed. This is due to the changed sign of the mean value of
$w_1$. And the HM prior seems favor more negative value of $w_1$
and stronger evolution of EOS.

\begin{figure}[htbp]
\begin{center}
\includegraphics[scale=0.5]{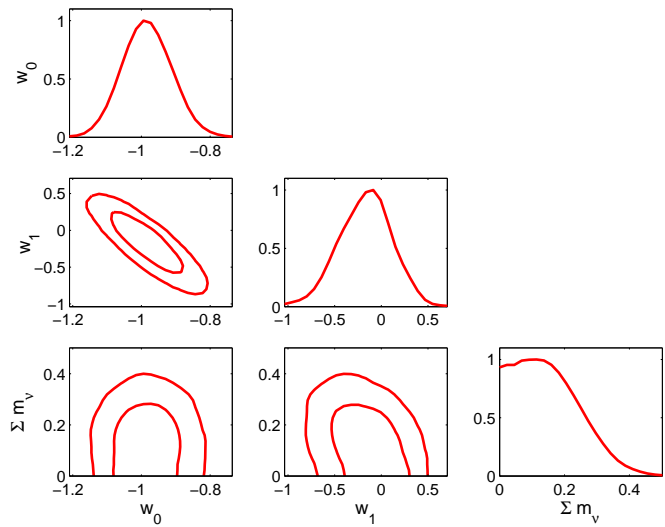}
\caption{Constraints on dark energy parameters
 $w_0$, $w_1$ and the total neutrino mass $\Sigma m_{\nu}$ from the Monte Carlo method.
 The two dimensional contours stand for $68\%$ and $95\%$ confidence levels \label{sim} .}
\end{center}
\end{figure}

In Ref.\cite{Hannestad:2005gj}, the author just used the constant
EOS to consider the correlation between EOS of dark energy and
neutrino mass. The critical point of our result is the correlation
among $w_0,~w_1$ and neutrino mass. From the table I we find there
\emph{exists} some correlation of ($w_0,~m_\nu$) and especially for
$w_1$ and $m_\nu$. More interestingly, we find using future high
quality data, the correlation between $w_0$ and $m_\nu$ nearly
vanishes while there is a strong and negative correlation between
the ``running" of dark energy EOS $w_1$ and the total neutrino mass
$\Sigma m_\nu$. Fig.(\ref{sim}) delineates the fitting result of
future simulation in more detail and we clearly see the strong
correlations between $w_1$ and neutrino mass. In
Ref.\cite{Ichikawa:2005hi}, the authors conclude $w_1$ is not
strongly correlated with neutrino mass basing on Fisher matrix
analysis. However, we have seen the distribution of $w_1$ and
neutrino mass is highly non-Gaussian and Fisher matrix technique is
not viable in such cases \cite{Perotto:2006rj}. Furthermore, they
didn't consider dark energy perturbation when EOS crosses $-1$. This
also give rise to great bias of determining neutrino mass limit if
neglecting time evolving of dark energy as illustrated in the table
I.

%Using the data of WMAP3+LSS+Ly$\alpha$+BAO, we find the $2\sigma$
%neutrino mass limit relax from $0.398eV$ (constant $w$) to
%$0.490eV$ (varying $w$), the relative error reaches more than
%$20\%$. If take $\Lambda$CDM model, the error will be greater than
%$70\%$!

The reason of the correlation between $\Sigma m_{\nu}$ and dark
energy is not hard to understand. The main correlation between dark
energy EOS and neutrino mass stems from the geometric feature of our
universe\cite{Hannestad:2005gj}. In addition, dynamical dark energy
will modify the time evolving potential wells which affects CMB
power spectra through the late time ISW effect. Dynamical dark
energy can leave imprints on CMB, LSS power spectrum and Hubble
diagram nonetheless these features can be mimicked by cosmic
neutrino to some extent\cite{Corasaniti:2002bw}. This is the source
of correlation.

%The effect is apparent when $w_1$ is large. Such strongly time
%evolving dark energy model can even be distinguished from
%$\Lambda$CDM model by CMB only .

%\begin{figure}[htbp]
%\begin{center}
%\includegraphics[scale=0.9]{merge.eps}
%\caption{Illustrative plots to show the correlation among the sum of
%neutrino mass $\Sigma m_{\nu}$ and dynamics of dark energy by
%studying their imprints on Hubble diagram, CMB TT power spectrum as
%well as matter power spectrum.
% \label{imprint}}
%\end{center}
%\end{figure}

\section{Conclusion}

In this paper we have done the global fitting including parameters
expressing time evolving dark energy, neutrino mass and other 6
cosmological parameters using different combination of currently
data as well as simulated data for future observation. We have seen
the interesting correlation between neutrino mass and EOS of time
evolving dark energy especially for $\Sigma m_{\nu}$ and $w_1$. The
correlation between the ``running" of dark energy EOS $w_1$ and the
total neutrino mass $\Sigma m_\nu$ is of great significance in
determining neutrino mass limit and revealing the nature of dark
energy by futuristic high precision observational data, such as
SNAP, PLANCK\footnote{http://sci.esa.int/planck/} and so forth.

{\bf Acknowledgements:} We acknowledge the use of the Legacy Archive
for Microwave Background Data Analysis (LAMBDA). Support for LAMBDA
is provided by the NASA Office of Space Science. Our MCMC chains
were finished in the Shuguang 4000A system of the Shanghai
Supercomputer Center(SSC). This work is supported in part by
National Natural Science Foundation of China under Grant Nos.
90303004, 10533010 and 19925523. We are indebted to Patrick Mcdonald
and Anze Slosar for clarifying correspondence on the fittings to the
Lyman $\alpha$ data. We thank Antony Lewis for technical support on
cosmocoffee\footnote{http://cosmocoffee.info}. We thank Bo Feng,
Hiranya Peiris, Mingzhe Li, Pei-Hong Gu, Hong Li and Xiao-Jun Bi for
helpful discussions and comments on the manuscript.

\end{document}